\documentclass[
	article,			
	12pt,				
	oneside,			
	a4paper,			
	english,			
	brazil,				
	sumario=tradicional
	]{abntex2}

\usepackage{lmodern}			
\usepackage[T1]{fontenc}		
\usepackage{indentfirst}		
\usepackage{nomencl} 			
\usepackage{color}	
\usepackage{url}			
\usepackage{graphicx}			
\usepackage{microtype} 			
\usepackage{amsmath}		
		
\usepackage[num]{abntex2cite}	

\newcommand\Real{\mbox{\textit{Re}}}  

\titulo{Drag force in wind tunnels: a new method}
\autor{
\hspace{-92pt}Paulo Victor Santos Souza\thanks{paulo.victor@ifrj.edu.br} \\
\hspace{-92pt}Instituto de Ciências Exatas, Universidade Federal Fluminense, Volta Redonda\\
\hspace{-92pt}Instituto Federal de Educação, Ciência e Tecnologia do Rio de Janeiro\\
 \hspace{-92pt} Volta Redonda, Rio de Janeiro, Brazil\\
\hspace{-92pt} Daniel Girardi \thanks{girardi1309@gmail.com}\\
\hspace{-92pt} Departamento de Ci\^encias Exatas e Licenciaturas, Campus Blumenau\\ 
 \hspace{-92pt}Universidade Federal de Santa Catarina,Blumenau, Santa Catarina, Brazil\\
\hspace{-92pt}Paulo Murilo Castro de Oliveira\thanks{pmco@if.uff.br}\\
\hspace{-92pt}Instituto de Física, Universidade Federal Fluminense\\
\hspace{-92pt} Niterói, Rio de Janeiro, Brazil\\
\hspace{-92pt}Instituto Mercosul de Estudos Avançados, Universidade Federal da Integração Latino Americana\\
\hspace{-92pt} Foz do Iguaçu, Paraná, Brasil
}
\local{Brazil}

\definecolor{blue}{RGB}{41,5,195}

\makeatletter
\hypersetup{
		pdftitle={\@title}, 
		pdfauthor={\@author},
    	pdfsubject={Modelo de artigo científico com abnTeX2},
	    pdfcreator={LaTeX with abnTeX2},
		pdfkeywords={abnt}{latex}{abntex}{abntex2}{atigo científico}, 
		colorlinks=true,       		
    	linkcolor=blue,          	
    	citecolor=blue,        		
    	filecolor=magenta,      		
		urlcolor=blue,
		bookmarksdepth=4
}
\makeatother

\setlrmarginsandblock{3cm}{3cm}{*}
\setulmarginsandblock{3cm}{3cm}{*}
\checkandfixthelayout

\SingleSpacing

\begin{document}

\selectlanguage{brazil}

\frenchspacing

\maketitle




\renewcommand{\resumoname}{Abstract}
\begin{resumoumacoluna}
	A rigid object of general shape is fixed inside a wind tunnel. The drag force exerted on it by the wind is determined by a new method based on simple basic Physics concepts, provided one has a solver, any solver, for the corresponding dynamic Navier-Stokes equation which determines the wind velocity field around the object. The method is completely general, but here we apply it to the traditional problem of a long cylinder perpendicular to the wind.
\begin{otherlanguage*}{english}

   \vspace{\onelineskip}
 
   \noindent
   \textbf{Keywords}: fluid dynamics, viscous flows, wind tunnels, drag force.
 \end{otherlanguage*}  
\end{resumoumacoluna}

\textual

\section{Introduction}
\label{introduction}

	Consider a rigid object facing the counter flux of a viscous fluid (air or water, for instance). The fluid velocity field around the object (which would be uniform in its absence) is distorted by its presence. This velocity field $\vec{v}(\vec{r})$ can be obtained as a function of time $t$ by solving the so-called Navier-Stokes dynamic equations \cite{Navier,Stokes}. These equations can be written in different forms, here we adopt a simple one

\begin{equation}\label{eq:NavierStokes}
\frac{\partial\vec\Omega}{\partial t} = \frac{1}{\Real}
\nabla^2 \vec\Omega - \vec\nabla\times(\vec\Omega\times\vec{v}),
\end{equation}

\noindent see for instance \cite{Feynman65}, where

\begin{equation}\label{eq:vorticity}
\vec\Omega = \vec\nabla \times \vec{v}
\end{equation}

\noindent is the vorticity field. Also, provided all speeds are small compared with the sound propagation speed in the same fluid, it can be considered incompressible, \textit{i.e.},

\begin{equation}\label{eq:continuity}
\vec\nabla\bullet\vec{v}=0.
\end{equation}

	The above equation (\ref{eq:NavierStokes}) is already expressed in adimensional units, considering the wind speed $V$ far from the object and some characteristic linear dimension of the rigid object such as the diameter $D$ of a cylinder, for instance. Together the characteristic fluid properties $\rho$, its uniform density, and $\mu$, its viscosity, these parameters can be condensed into a single adimensional Reynolds number \cite{Reynolds}

\begin{equation}\label{Reynolds}
\Real= \frac{\rho V D}{\mu}.
\end{equation}

\noindent This way, one can consider $V$ in equation (\ref{eq:NavierStokes}) as the speed unit, and $D$ as the length unit. The corresponding time unit is $D/V$.

	These 4 above equations determine completely the velocity or vorticity fields around the object, as functions of time, given the proper boundary and initial conditions. Normally, the boundary condition is $v=1$ far from the object along some fixed direction, and $v=0$ at the object surface. The initial condition may be (as here) the wind tunnel initially switched off and suddenly switched on with a fixed value $V$ (or $\Real$, impulsive switching). In this case, one has first to solve the time independent Stokes equation $\nabla^2 \vec\Omega = 0$, obtained by taking $\Real\to 0$ in equation (\ref{eq:NavierStokes}), in order to obtain the initial fields $\vec{v}(\vec{r})$ and $\vec\Omega(\vec{r})$ at $t=0$. By neglecting this care with the initial fields, one obtains spurious transient fields just after $t=0$, while the steady-state final regime is not reached \cite{Cruchaga02}.

	What is not explicit in the above equations is how to determine the drag force the wind exerts on the fixed object. The traditional way to do this is by first determining the pressure distribution and the shear stresses on the object surface and then integrating them. However, these distributions are often extremely difficult to obtain, either experimentally or theoretically \cite{Munson90}. Alternatively, one can obtain the drag force just through the knowledge of the fluid velocity field around the obstacle. Also in this case, an integration over the surface of the object that involves the gradient of this field on the surface is necessary \cite{Noca99,Tan05}. If the method adopted to solve the above equations is a numerical one on a discrete grid, the determination of the quoted gradient requires a very fine grid near the surface, which takes a large computational effort. That is why normally researchers adopt  non-uniform grids, very fine only near the object surface, when the aim is to determine the drag force. This approach unnecessarily complicates the mathematical discretization procedures needed to translate the continuous derivatives into finite differences, besides the further computer effort fine grids require.

	The current text introduces an alternative method to determine the drag force replacing the surface integration by a volume integration, which solves both problems commented at the end of last paragraph. The method is completely general and based only on simple basic Physics concepts. It is described below, and its validity mathematically demonstrated at the Appendix (although this mathematical demonstration is completely unnecessary, the conceptual arguments below are enough).

	Consider one has a solver for the Navier Stokes equations above, providing the field $\vec{v}$ in all space at time $t + \delta t$ from the knowledge of the previous field at time $t$, where $\delta t$ is some small time interval. Any solver can be used, the accuracy of the method now described depends only on the accuracy provided by this solver. 
Imagine one replaces the rigid object by an extra portion of fluid at time $t$. This extra portion of fluid is also at rest in the same way as the removed object. But it would move from time $t$ on. A small fluid velocity would appear at each point inside the volume formerly occupied by the object, at $t + \delta t$. By integrating these velocities inside the quoted volume and multiplying the result by the density of the fluid, one obtains the total linear momentum which would be transferred from the fluid to the object, a vector. This would-be transferred momentum is compensated by the mechanical device keeping the object at rest inside the wind tunnel. Finally, dividing this total linear momentum by $\delta t$, one obtains the force exerted by the original fluid on the object.

	In short: I) One has the velocity field at time $t$ around the object, with all other velocities null inside the volume formerly occupied by the object, now occupied by the static fluid replacing it; II) One uses the Navier Stokes equations solver in order to obtain the same field at $t + \delta t$; III) Non-null, small velocities appear inside the quoted volume; IV) One integrates these penetrating small velocities over the quoted volume, multiply the result by the fluid density and divide it by $\delta t$. 

	It is worth mention why the Newtonian momentum transfer procedure is enough in our case, no convective contributions are needed. There are two different descriptions to treat continuous fluid movements. The Lagrangian description takes the control volume of fluid moving with the fluid, the portion of fluid inside this volume is always the same, nothing entering or leaving this control volume. In this description, one can directly apply Newton's law. The Eulerian description takes a fixed and static control volume, the fluid passing through it. The relation between these two descriptions is made through the so-called convective derivatives. The material derivative $(D/Dt)$ applicable to the Eulerian description is obtained by adding the quoted convective derivative $(v_x \partial/\partial x + v_y \partial/\partial y + v_z \partial/\partial z)$ to the regular time derivative $(d/dt)$ applicable to the Lagrangian description \cite{Schneiderbauer14}. This extra term is just the origin for convective corrections over merely Newtonian inertial forces. However, in our particular case, by replacing the object by static fluid, our control volume is that of the object itself. Because the fluid is static at this precise time $t$, the Lagrangian and Eulerian volume controls are exactly the same, there are no translations between them $(\partial/\partial x = \partial/\partial y = \partial/\partial z = 0)$, therefore the convective derivative of any quantity vanishes. That is why convective terms don't enter in our momentum transfer formulation. This also applies to the case of a rotating long cylinder perpendicular to the wind  treated later, the convective derivative is also null in this case.

	This text is organized as follows. In next section we show the results for the drag force on a long static cylinder, perpendicular to the wind, inside a wind tunnel initially switched off and suddenly switched on in $t=0$ with a Reynolds number $\Real= 1,000$. After some transient time, the steady state drag force reaches a value to be compared with the experimentally known value of the drag coefficient $C_D \approx 1.0$ \cite{Wieselsberger1921,Finn1953,Tritton1959,Roshko1961,Jayaweera1965,Kato1983}. The agreement with this experimental value is excellent (see figure \ref{forcaestatico} discussed later). Before that, however, starting at $t=0$ the drag force increases and reaches larger values, decreasing a little bit up to stabilizing at the steady state. This behavior is the same observed in experiments as \cite{Sarpkaya66,Taneda72,Sarpkaya78,Nomura00,Takeuchi08,Shirato09,Takeuchi10}, in particular that of a small and light falling ball which surprisingly reaches a larger speed than its final limit speed during the transient time \cite{Oliveira10}. The explanation of this curious phenomenon resides on the gradual formation of the von K\'arm\'an street of vortices behind the ball: While it is not yet completely formed, the drag force is smaller than its own steady state value, allowing larger fall speeds; After the street of vortices is completely formed, the drag force surpasses the ball weight and the ball breaks; At the end, the drag force decreases a little bit becoming equal to the ball weight, and hereafter the limit speed remains constant. In the following section, a rotating cylinder is treated. Besides the drag, the lift coefficient due to the Magnus effect is also determined for various angular velocities of the cylinder. In all cases, the formation of vortices behind the cylinder is dynamically shown. Finally, the last section resumes the whole work.
\newpage
	\section{Drag force on a long static cylinder}
	
	In this section, one describes the flow over a long static cylinder immersed in a wind tunnel initially switched off and suddenly switched on in $t=0$ with a Reynolds number $\Real = 1,000$\footnote{ One uses the method of successive relaxations to solve the Navier-Stokes equations \cite{Oliveira12}. However, any other method may be used.} and, then, one applies the method described in the previous section to determine the behavior of the drag force on this cylinder.
	
\begin{figure}[h!]
	\centerline{\includegraphics[width=350pt]{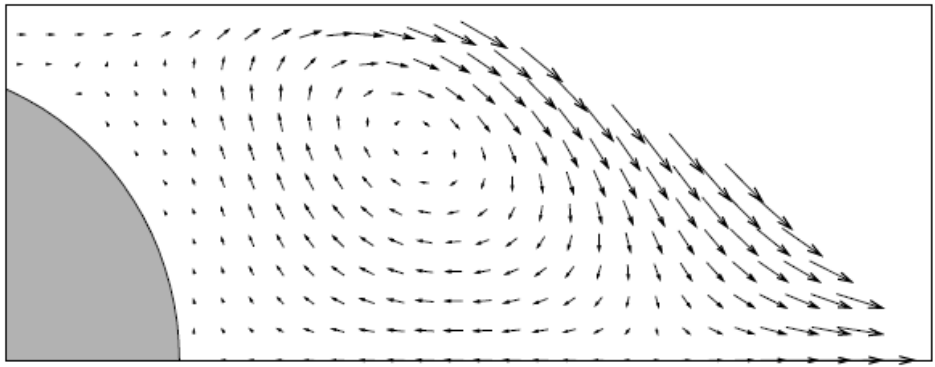}}
	\caption{Starting from the Stokes configuration, the wind is switched on at $t = 0$ with a fixed Reynolds number $\Real = 1,000$. The figure show the result at $t = 1$, when a fluid element far from the cylinder has already traveled one diameter.  Just a close up of the region behind the cylinder, above the $X$ axis, is shown. The other symmetric vortex below is not shown.
}
	\label{evol1}
\end{figure}

\begin{figure}[h!]
	\centerline{\includegraphics[width=350pt]{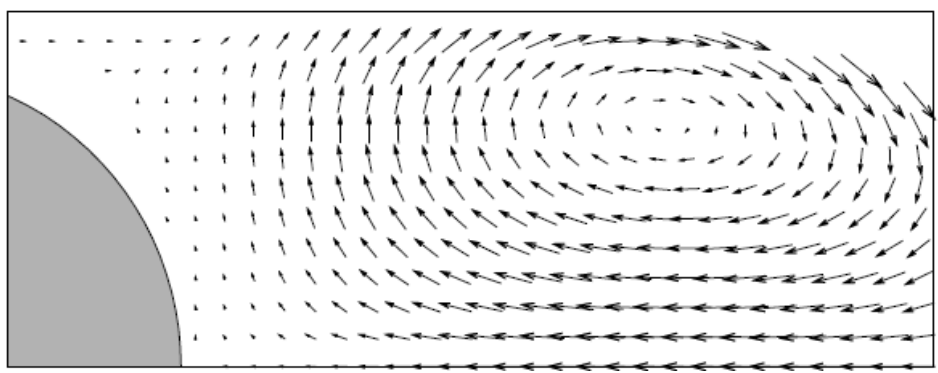}}
	\caption{Later than figure \ref{evol1}, $t = 2$. Vortices are still symmetric.}
	\label{evol2}
\end{figure}

	Starting from the Stokes configuration, the wind is switched on at $t = 0$ with a fixed Reynolds number $\Real = 1,000$. After a transient time, two symmetric vortices are formed behind the cylinder, a close up of one of them is shown in figure \ref{evol1}. Some time later, they move a little bit downstream and start to become stretched along the $X$ direction, as shown in figure \ref{evol2}.

	After that, one vortex bifurcates and becomes a pair of vortices running in the same sense and repelling each other along the $X$ direction. Later on, so does the other vortex. Only then the aforementioned von K\'arm\'an street begins to be formed: one vortex running clockwise slowly goes away downstream, followed by another running counterclockwise, and so on. 	
	
	A completely description about the flow on a static cylinder can be found in reference \cite{Oliveira12}. There, the velocity field is considered two dimensional on the plane perpendicular to the cylinder, a traditionally adopted good approximation for these not-so-high Reynolds numbers. The external boundary condition is a $10D \times 5D$ rectangle, measured in terms of the cylinder diameter $D$. Outside this rectangle the velocity field $V$ is considered uniform along the direction of its two parallel largest edges. The cylinder center is located inside this rectangle at a distance $3D$ from the incoming wind edge and $2.5D$ from both largest edges. The discretization numerical grid covering this rectangle has $400 \times 200$ pixels, therefore each pixel is a small $\frac{1}{40}D \times \frac{1}{40}D$ square. The discretized time interval adopted to solve the Navier-Stokes equations is $\frac{D}{400V}$, the time wind spends to travel one tenth of a pixel.

\begin{figure}[h!]
	\centerline{\includegraphics{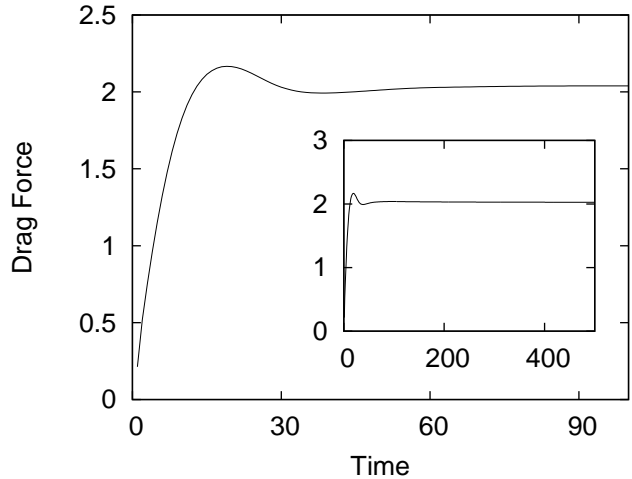}}
	\caption{Drag force on a static cylinder in the wind tunnel initially switched off. It is turned on at $t=0$  with a constant Reynolds number ${\Real}=  1,000$, which defines the speed of the wind. The force in the  direction perpendicular to the wind flow is negligible. The dimensionless drag force is plotted in terms of the so-called drag coefficient. During one time unit the wind travels one cylinder diameter. The force increases, reaches a maximum value, decreases and finally stabilizes, corresponding to a dynamic situation where successive vortices appear continuously, slowly going away, one running clockwise, the next running counterclockwise and so on. The long von K\'arm\'an street is then formed. Inset shows the same in a longer time scale.}
	\label{forcaestatico}
\end{figure}

	So, one may now apply the previously described method to determine the behavior of the drag force on the static cylinder. Figure \ref{forcaestatico} shows the drag force as a function of time.\footnote{ The time $\delta t$ adopted is the same as the discrete time interval used to solve the Navier-Stokes equation, i.e. the time the wind takes to transverse 0.1 grid pixel.}
	
Analysis of the graph shows that the drag force on the static cylinder increases, reaches a maximum value, decreases, and stabilizes. As expected, the same behavior of the falling ball \cite{Oliveira10} is found for this large Reynolds number, for which the long von K\'arm\'an street of successive vortices is present. On the other hand, for small values of $\Real$, about a few dozens, for which instead of the long street periodically fed with new vortices only two vortices appear and stabilize behind the cylinder, the force does not present a maximum value before stabilization. Also, in this case the drag force is much smaller.

The existence of a transient period during which the drag force reaches a value larger than its steady state is confirmed by various experimental studies \cite{Sarpkaya66,Taneda72,Sarpkaya78,Nomura00,Takeuchi08,Shirato09,Takeuchi10}. Moreover, our results are in accordance with known experimental data. A comparison between the results obtained by our method and experimental data is made through the drag coefficient \cite{Munson90},
\begin{equation}\label{cd}
C_D = \frac{F_{drag}}{\frac{1}{2}\hspace{2pt} \rho\hspace{2pt} V^2\hspace{2pt} D\hspace{2pt}l},
\end{equation}
where $F_{drag}$ is the parallel-to-the-wind component of the drag force, $\rho$ is the density of the fluid, $V$ is the wind's speed, $D$ is the diameter of the cylinder  and $l$ is the length of the cylinder. Drag coefficients are measured experimentally since the early twentieth century. The drag coefficient measured at steady state for a static cylinder was measured for the first time (within the knowledge of the authors) in 1921 \cite{Wieselsberger1921}. After that, several other similar experiments were carried out, and today it is a known fact that for $Re=1,000$, $C_D \approx 1.0$ \cite{Finn1953,Tritton1959,Roshko1961,Jayaweera1965,Kato1983}, whereas our method provides $C_D = 1.01$.
\section{Drag force on a long rotating cylinder}

In a different version of the problem, the cylinder rotates with constant angular speed $\omega$. Now, the fluid velocities on the surface of the cylinder are not null, but are equal to the speed at the surface of the cylinder, i.e., the fluid rotates with the cylinder. The angular speed $\omega$ is measured in units such that $\omega=1.0$ corresponds to the magnitude of the velocities on the surface of the cylinder equal to the speed of the wind. 

\begin{figure}[h!]
		\includegraphics{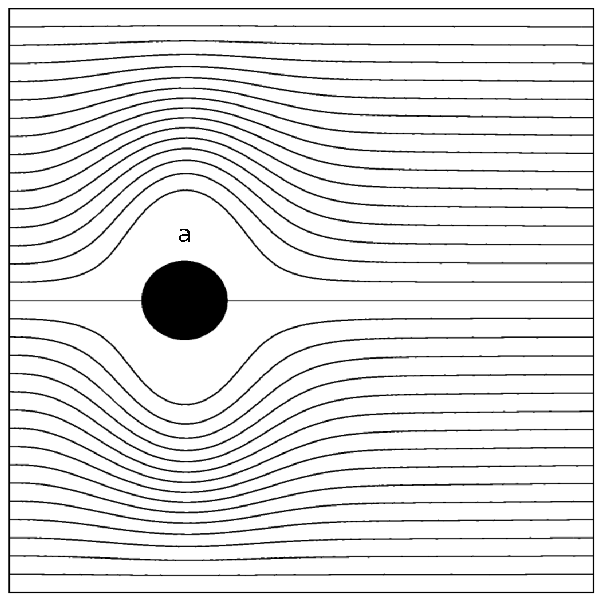} 
		\hspace{1pt}
		\includegraphics{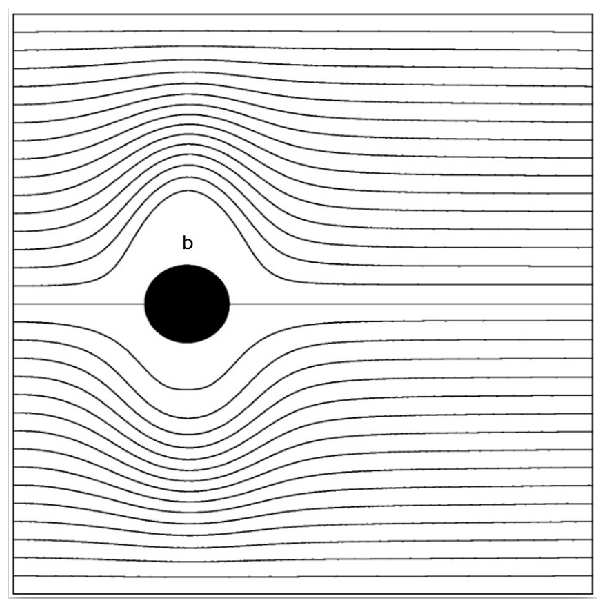} 
		\caption{Streamlines for Stokes's configuration 
		(${\Real} \to 0$), when the cylinder rotates clockwise with (a) $\omega = 0.1$ and (b) $\omega = 0.5$. The rotation of the cylinder results in the breakdown of the axial symmetry. As the angular speed increases, the lines are deformed, as expected. The wind goes from left to right.}
		\label{streamspin}
\end{figure}

Let's start with the Stokes laminar limit $(\Real \rightarrow$ $0)$ for which the Navier-Stokes equation reduces to the Laplace equation, $\nabla^2 \vec{\Omega} = 0$. The resulting streamlines are shown in figure \ref{streamspin}.

\begin{figure}[h!]
\begin{center}
	\includegraphics{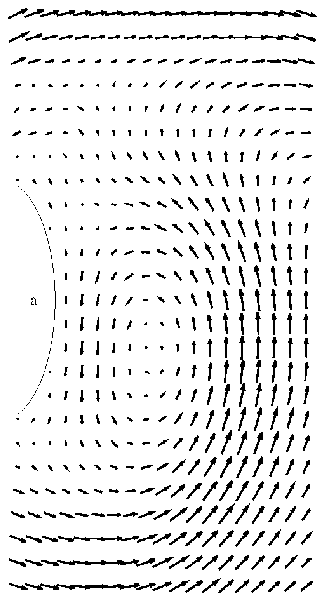} 
	\hspace{10pt}
	\includegraphics{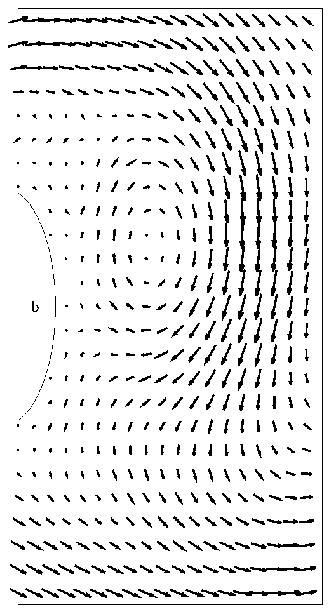} 
	\hspace{10pt}
	\includegraphics{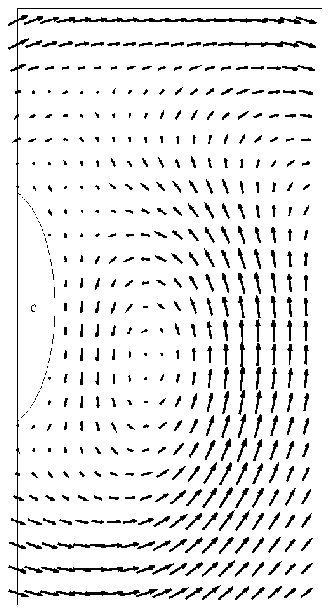}
	\caption{Starting with Stokes's laminar configuration for the clockwise rotating cylinder, the wind is switched on at $t = 0$, with a fixed Reynolds number $\Real = $ 1,000 and $\omega = 0.5$. After a transient time, the velocity field changes continually and periodically. Figure (a) shows a vortex that has just formed behind the cylinder, rotating in the counterclockwise direction. It then moves away to the right. Some time later, another vortex is formed behind the cylinder, rotating in the clockwise direction, figure (b). Later, figure (c), the velocity field returns to the configuration  observed in figure (a). The whole process is periodically repeated. At this fine scale close to the cylinder back side, the behavior is the same observed in the case of a fixed cylinder. However, behind this region there is a long von K\'arm\'an street of already previously formed vortices, now bended downwards in this slowly rotating case considered here.}
		\label{reynoldsspin}
		\end{center}
	\end{figure}

When the original Navier-Stokes equation is solved for a fixed and large Reynolds number, say $\Real=1,000$, the solution shows that, after a transient time, as in the case of the static cylinder, the system cyclically stabilizes: a vortex rotating in the counterclockwise direction appears in the region behind the cylinder and below the horizontal axis formed by its diameter, see figure \ref{reynoldsspin}(a); then, this vortex grows and slowly moves away downstream; after some time, a second vortex appears in the region above the horizontal axis, rotating in the clockwise direction, see figure \ref{reynoldsspin}(b); this vortex also grows and moves away; later, a third vortex rotating in the counterclockwise direction begins to form in the region below the horizontal axis, see figure \ref{reynoldsspin}(c); and this process repeats again and again.

	The method  to determine the behavior of the drag force, presented in the first section, may now be applied to the case of the rotating cylinder with one difference: the volume occupied by the cylinder is initially filled by fluid moving exactly as the cylinder, \textit{i.e.}, rotating as a rigid body at time $t$. Then, starting from this configuration, the flow evolves up to $t+{\mathrm \delta}t$.

\begin{figure}[h!]
	\begin{center}
		\includegraphics{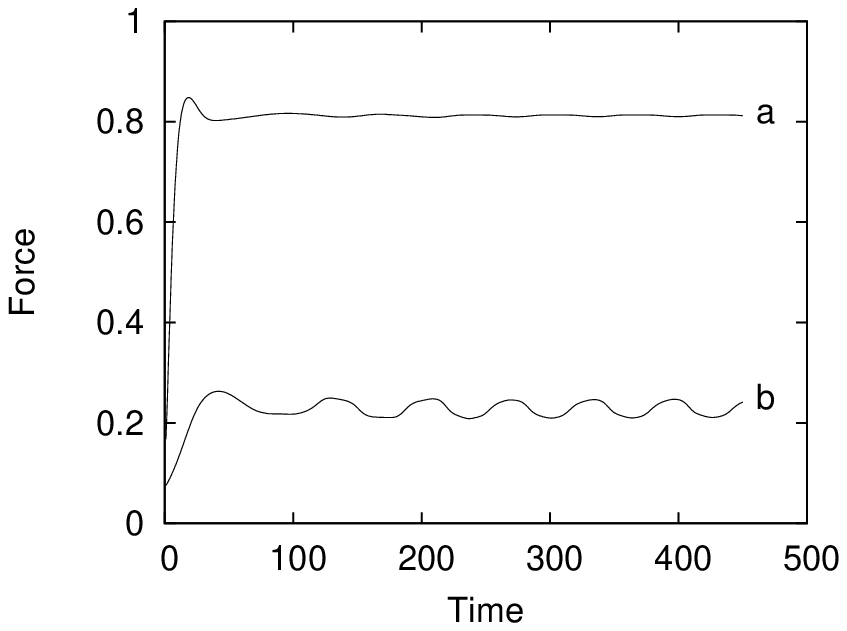}
	    \caption{The drag force on a rotating cylinder inside the wind tunnel with Reynolds number $ \Real = $ 1,000, and $\omega = 0.1$. The drag coefficient (a) shows the same behavior as before, in the case of fixed cylinder. During the transient stage, the lift coefficient (b) also increases, reaches a maximum, and decreases. However, this component does not stabilize. Instead, it oscillates around a non null mean value. The oscillation is due to the alternate formation of counterclockwise and clockwise vortices, being a signature of the von K\'arm\'an street. For small values of $\Real$, up to a few dozens, both components of the drag force don't oscillate, a signature of the absence of successive vortices.}
	    \label{forcaspin}
	    \end{center}
\end{figure}

	In this case, due to rotation, the force on the cylinder has a non zero component in the direction transverse to the wind. The components of the force on a cylinder that rotates with $\omega =0.1$ are shown in figure \ref{forcaspin}, where the forces are expressed in terms of drag and lift coefficients. As the drag coefficient, the lift coefficient is an adimensional coefficient widely used in fluid dynamics and defined as \cite{Munson90}
\begin{equation}\label{cl}
C_L = \frac{F_{lift}}{\frac{1}{2}\hspace{2pt} \rho\hspace{2pt} V^2\hspace{2pt}D\hspace{2pt}l},
\end{equation}
where $F_{lift}$ is the transverse-to-the-wind component of the drag force.
The emergence of the transverse non-zero component is due to an asymmetry in the boundary layer
	\footnote{ The boundary-layer results from the adherence of the air
		molecules to the object, say a cylinder. Due to 
		viscosity, the adherence is partially transmitted to the molecules
		situated farther from the cylinder; this defines a region that moves
		with the cylinder, called boundary-layer or layer of Prandtl \cite{Munson90}.}
	separation, triggered first near the bottom of the  cylinder (as in figures \ref{streamspin} and \ref{reynoldsspin}, the cylinder is considered to rotate clockwise). This is a clear manifestation of the well known Magnus effect \cite{Munson90}.

\begin{figure}[h!]
	\begin{center}
		\includegraphics{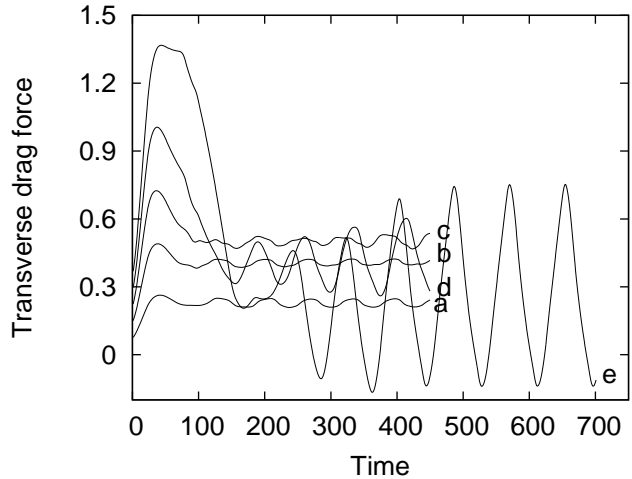}
	    \caption{Lift coefficients as a function of time, for various small angular speeds of the rotating cylinder with $\Real= 1,000$: (a) $\omega$ = 0.1; (b) $\omega$ = 0.2; (c) $\omega$ = 0.3; (d) $\omega$ = 0.4; and (e) $\omega$ = 0.5. After the transient stage, the Magnus force oscillates around a non null average value.}
	\label{magnusomegas}
	\end{center}
\end{figure}

	As a net result, the von K\'arm\'an street as a whole is deflected downwards, explaining the non null average in figure \ref{forcaspin}, curve (b), and figure \ref{magnusomegas}, where lift coefficients as a function of time for various small angular speeds of the rotating cylinder are shown. The oscillations are due to the appearance of successive vortices spinning in alternate senses, forming the street. Close to the cylinder, a counterclockwise vortex pushes it downwards, subtracting an extra force from the average value, see figure \ref{reynoldsspin}(a) corresponding to a minimum of the Magnus force in figure \ref{forcaspin}, curve (b). A clockwise vortex, see figure \ref{reynoldsspin}(b), pushes the cylinder upwards, adding an extra force to the average, corresponding to a maximum of figure \ref{forcaspin}, curve (b). It is interesting to observe that the drag coefficient also oscillates, hardly visible in figure \ref{forcaspin}, curve (a) (see also \cite{Cruchaga02, Lu02}).

\begin{figure}[h!]
		\begin{center}
		\includegraphics{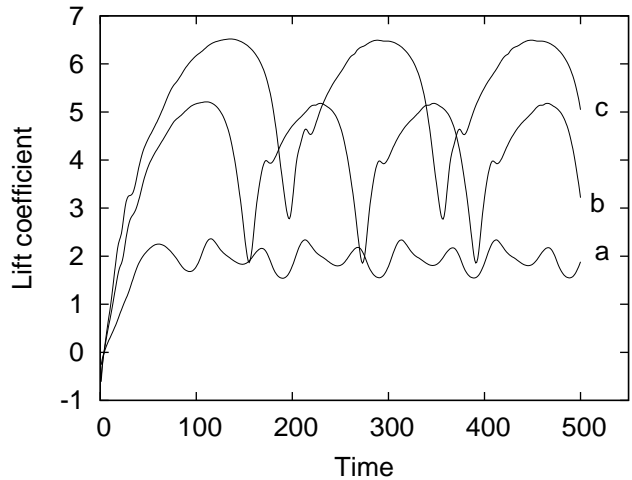} 

	\caption{Lift coefficients as functions of time, for various angular speeds of the rotating cylinder with $\Real= 1,000$: (a) $\omega$ = 1.0; (b) $\omega$ = 2.0; and (c) $\omega$ = 2.4. When $\omega \approx 2.0$, the high-frequency oscillations are replaced by much-smaller-frequency oscillations, due to the absence of the von K\'arm\'an street.} 
	\label{high}
	\end{center}
\end{figure}

For larger angular speeds $ (\omega \approx 2.0) $, it is known that the street of vortices tends to disappear \cite{Chew95,Chen93, Stojkovic02, Mittal03, Kalinin11, Rao13, Rao14}. Figure \ref{high} shows the behavior of the lift coefficient for some values of $\omega$ larger than before. Our results indicate that, for angular speeds larger than $ \omega \approx 1.0 $, the high-frequency oscillations related to the sequence of successive vortices are attenuated as the angular speeds increases and tend to disappear for angular speed values close to  $\omega \approx 2.0$, replaced by much-smaller-frequency oscillations.

\newpage
\section{Final Remarks}
In this paper, we introduce a new method to determine the dynamic behavior of the drag force acting on an object inside a wind tunnel. 

Using this method, it was revealed that, in the case of a static cylinder, the drag coefficient increases, reaches a maximum value, decreases and stabilizes. This behavior agrees with the experimental results presented in various references \cite{Sarpkaya66,Taneda72,Sarpkaya78,Nomura00,Takeuchi08,Shirato09,Takeuchi10,Oliveira10}. Moreover, the drag forces we found are in excellent agreement with known experimental data. These two facts attest to the validity and effectiveness of the method.

In the case of a slowly rotating cylinder, the drag coefficient behaves in the same way as in the case of the static cylinder. However, the lift coefficient is not null: it grows, reaches a maximum, decreases and fluctuates around a non null value. This is a clear manifestation of the well known Magnus effect \cite{Munson90}. Moreover, also in this case, the method is even capable of identifying a transition scheme where the von K\'arm\'am street of vortices disappears as the angular speed increases.

The method, as we hope it has been clear, is simple, because it requires no knowledge about the pressure distribution and the shear stresses or the gradient of the fluid velocity field on the object surface. No approximation is 
        used (except the numerical grid discretization) and no convective correction is 
        necessary. The method is also applicable to objects of general shape.
\section{Acknowledgments}
This work is partially supported by  the Brazilian agencies CAPES and CNPq. The authors thank P.M.C. Dias for her comments on this work.


\section*{References}

\bibliography{mybibfile}
\section{Appendix}
\label{appendix}

	In this appendix, we demonstrate mathematically the validity of the method described in the first section. 
	
	The object (cylinder or any other) inside the wind tunnel is fixed there by some 
mechanical device exerting a force $-\vec{F}$ on it in order to counterbalance the 
fluid drag force $\vec{F}$. At time $t$ one can imagine the object replaced by fluid,
also at rest. Now, the extra portion of fluid would move. A fluid element ${\mathrm d}^3r$
inside the former object volume will suffer a force ${\mathrm d}\vec{f}(\vec{r})$ exerted by 
the rest of fluid around it, and thus would acquire a velocity $\delta\vec{v}$ after a 
small time interval $\delta t$. Newton's law 

\begin{equation}\label{NewtonLaw}
{\mathrm d}\vec{f}(\vec{r}) = \rho\, {\mathrm d}^3r\, \frac{\delta\vec{v}}{\delta t},
\end{equation}

\noindent can be applied, where $\rho$ is the fluid density.

	In order to avoid this extra movement keeping the extra portion of fluid at rest,
some ``magical'' device must exert a second force $-{\mathrm d}\vec{f}(\vec{r})$ on the
fluid element ${\mathrm d}^3r$. Considering the whole extra volume also at rest, one can
determine the resultant force this ``magical'' device must exert on the whole volume
$V_{\mathrm ob}$ formerly occupied by the object

\begin{equation}\label{reaction}
-\vec{F} = - \int_{V_{\mathrm ob}}\, {\mathrm d}\vec{f}(\vec{r}) = 
- \rho\, \frac{\int_{V_{\mathrm ob}}\, {\mathrm d}^3r\, \delta\vec{v}}{\delta t}
\end{equation}

	Now, this ``magical'' device becomes not magical at all, it is just the same already quoted mechanical device fixing the object inside the wind tunnel. Therefore, the drag force exerted by the fluid on the object is

\begin{equation}\label{drag}
\vec{F} = \rho\, \frac{\int_{V_{\mathrm ob}}\, {\mathrm d}^3r\, \delta\vec{v}}{\delta t},
\end{equation}

\noindent equation which defines our method.

	Note that the imagined replacement of the object by static fluid is made only at the time $t$, no velocity differences appear. In other words, the above equation
is exactly the result of Newton's law as applied to the object. No error, no approximation. Only one source of inaccuracies may appear: the adopted time interval $\delta t$ is not infinitesimal, it must be chosen small enough in the same standards as the also finite time interval adopted by the numerical method used to solve the Navier-Stokes equation. Therefore, the possible inaccuracies of our method to calculate the drag force come exclusively from the numerical method adopted to solve the Navier-Stokes equation. These possible inaccuracies are of course controllable.

\end{document}